# Real-Time COVID-19 Diagnosis from X-Ray Images Using Deep CNN and Extreme Learning Machines Stabilized by Chimp Optimization Algorithm

Hu Tianqing[1], Mohammad Khishe[2*], Mokhtar Mohammadi[3], Gholam-Reza Parvizi[4], Sarkhel H. Taher Karim[5], Tarik A. Rashid[6],

[1]College of Computer Science and Technology, Henan Polytechnic University, Jiaozuo City, Henan Province, China.

[2*]Corresponding Author, Department of Electronic Engineering Imam Khomeini Marine Science University, Nowshahr, Iran, https://orcid.org/0000-0002-1024-8822, m_khishe@alumni.iust.ac.ir

[3]Department of Information Technology, Lebanese French University, Erbil, KRG, Iraq.

[4]Faculty of Foreign Languages, University of Isfahan, Isfahan, Iran.

[5]Computer Department, College of Science, University of Halabja, Halabja, Iraq.

[6]Computer Science and Engineering Department, University of Kurdistan Hewler, Erbil, KRG, Iraq.

**Abstract**

Real-time detection of COVID-19 using radiological images has gained priority due to the increasing demand for fast diagnosis of COVID-19 cases. This paper introduces a novel two-phase approach for classifying chest X-ray images. Deep Learning (DL) methods fail to cover these aspects since training and fine-tuning the model's parameters consume much time. In this approach, the first phase comes to train a deep CNN working as a feature extractor, and the second phase comes to use Extreme Learning Machines (ELMs) for real-time detection. The main drawback of ELMs is to meet the need of a large number of hidden-layer nodes to gain a reliable and accurate detector in applying image processing since the detective performance remarkably depends on the setting of initial weights and biases. Therefore, this paper uses Chimp Optimization Algorithm (ChOA) to improve results and increase the reliability of the network while maintaining real-time capability. The designed detector is to be benchmarked on the *COVID-Xray-5k* and *COVIDetectioNet* datasets, and the results are verified by comparing it with the classic DCNN, Genetic Algorithm optimized ELM (GA-ELM), Cuckoo Search optimized ELM (CS-ELM), and Whale Optimization Algorithm optimized ELM (WOA-ELM). The proposed approach outperforms other comparative benchmarks with 98.25% and 99.11% as ultimate accuracy on the *COVID-Xray-5k* and *COVIDetectioNet* datasets, respectively, and it led relative error to reduce as the amount of 1.75% and 1.01% as compared to a convolutional CNN. More importantly, the time needed for training deep ChOA-ELM is only 0.9474 milliseconds, and the overall testing time for 3100 images is 2.937 seconds.

## 1. Introduction

The early detection of Coronavirus has become a challenge for scientists due to the limited access to treatment and vaccines around the world. Polymerase Chain Reaction (PCR) tests have been introduced as one of the primary methods for detecting COVID-19 [1]. Nevertheless, this test is demanding and time-consuming; contrary, X-ray images are vastly accessible, and their scans are comparatively low in cost [2-5].

The need for designing an accurate and real-time detector has become increasingly necessitated. Considering the significant capabilities of Deep Learning (DL) in cases [6-8], we propose to apply DCNN as a COVID-19 detector. A few research studies have been conducted since the beginning of the year 2020, having attempted to develop methods for identifying patients affected by the pandemic via DCNN [9,10]. Although the significant features of DL enable it to solve different learning tasks, it is difficult to train it [11-13]. Some instances of successful methods for training Deep Learning (DL) are GD [14], Conjugate Gradient (CG) [15], Hessian-Free Optimization (HFO) algorithm [16,17], and Krylov Subspace Descent (KSD) [18].

Although stochastic GD training methods are simple in structure and rapid in process, they demand numerous manual parameters tuning to make them optimal [19,20]. Additionally, their strategies are inherently sequential; therefore, keeping their paces with Graphics Processing Units (GPU) is much demanding [21,22]; Although CG is stable for training, yet it is almost slow and needs multiple CPUs and a large number of RAM's resources [23-25].

Deep auto-encoder has used HFO to train the weights [16], being more efficient in pre-training and fine-tuning deep auto-encoders than the model proposed by Hinton and Salakhutdinov [17]. On the other side, KSD is more straightforward and stronger than HFO; it is additionally proven that KSD better classifies and optimizes than HFO. However, KSD requires more capacity than HFO [7].

Reference [26] proposes a Progressive Unsupervised Learning (PUL) approach to transfer pre-trained deep DCNN. This method is simple to conduct, and it can be viewed as a useful baseline for unsupervised-feature learning. As clustering results can be very noisy, this method adds a selecting operation between the clustering and the fine-tuning phases.

An automatic DCNN architecture design method that uses genetic algorithms is proposed in [27] to optimize the image classification problems. The proposed algorithm's main feature is related to its automatic characteristic, meaning that users do not need any knowledge of the DCNN's structure. However, this method's major drawback is that the GA's chromosomes become too large within large DCNNs getting the algorithm slowed down.

The mentioned methods are, at least, throughout the training phase time-consuming. Therefore, it takes hours for users to obtain feedback in advance if the selected model for detecting works in the intended case. Additionally, self-learning X-ray image detection that trains based on the user's feedback progressively may not have the right user experience since it takes a long time until the model enhances while operating with it [28-30]. A challenging point is an approach for X-ray image detection being efficient both in testing and in training phases.

This study proposes to use ELM [31] yet with a fully connected layer to provide a real-time training phase. In the two-phase proposed approach, we compound the automatic feature of deep CNNs learning with efficient ELMs to tackle the mentioned shortcomings, i.e., manual feature extraction and training time extension.

Consequently, the first phase is the deep CNN's training considered as an automatic feature extractor. In the second phase, ELM will be replaced by a fully connected layer for designing a real-time classifier.

The ELM's origin is based on the Random Vector Functional Link (RVFL) [32-34], leading to ultra-fast learning and significant-generalization capability. Previous surveys show that ELM has widely been used in many engineering applications [35-38]. Although different types of ELM [39-41] are now accessible for detecting image and classifying problems, these problems, including the need for many hidden nodes to better generalization and the choice of activation functions, remain intact. Besides, ELM's stochastic nature brings about an extra-uncertainty problem, particularly for high-dimensional image processing systems [42,43].

The ELM-based models randomly select the input weights and hidden biases from which the output weights are calculated during this procedure; ELMs attempt to minimize the training error and identify the smallest output weights' norm. Due to the stochastic choice of the input weights and biases in ELM, the output matrix may not indicate full column rank; Instead, it leads to ill-conditioned matrices to system producing non-optimal solutions [44]. Consequently, we use a novel meta-heuristic algorithm called Chimp Optimization Algorithm(ChOA) [45,46] to improve ELM conditioning and ensure optimal solutions. To Sum up, we propose to use ELM rather than the last conventional fully-connected layer in deep CNN to have both a real-time training phase and a real-time testing phase. It is necessary to note that traditional ELM suffers from ill-conditioning and uncertainty, which leads to proposing ChOA [45] maintaining real-time structure and detecting with high accuracy.

The remainder of this paper is structured as follows. Section 2 reviews background materials; Section 3 introduces the proposed scheme; Section 4 presents simulation, discussion, and results. Finally, conclusions are brought in Section 5.

## 2. Background and Materials.

This section represents the background knowledge, including the DCCN, ELM, ChOA, and COVID-19 datasets.

### 2.1 Deep Convolution Neural Network

Deep learning models have rapidly become a methodology for analyzing X-ray images [47-49]. As shown in [50,51], the most successful type of deep learning model for X-ray image analysis to date is CNN. CNNs consist of many layers that transform their input with convolution filters of a small extent. Various variants of CNNs have been proposed, such as LeNet-5 [52], AlexNet [53], ZFNet [54], GoogLeNet [55], VGGNet [56], ResNet [57], and etc. Given that the proposed model is supposed to evolve by metaheuristic algorithm, large networks can contribute to high computational cost since the evolutionary process is prolonged [58].

Having a big network can lead to overfitting [59]. LeNet is simple yet effective for grayscale and black and white images like chest X-ray images. Because of the limitations mentioned earlier, we propose LeNet as the primary classifier, to reduce the structural complexity and increase the chance of real-time processing. LeNet is the simplest type of CNNs introduced by Yann Le-Cun in the late 1990s, broadly considered as the first set of true CNNs [52]. Table 1 represents the details of the LeNet-5 architecture. These concepts can be arranged in classes of two layers, including sub-sampling layers and convolution layers. As shown in Fig. 1, the processing layers comprise three convolution layers that are located between sub-sampling layers, organized as feature maps. The final output layers are three fully connected layers.

Structurally viewed, any Feature Maps (FMs) is the outcome of a convolution from the previous layer's maps by its corresponding kernel and a linear filter. The weights $w^k$ and the adding bias $b_k$ generate the $k_{th}$ (FM) $FM_{ij}^k$ using the $tanh$ function as in Eq. (1).

$$FM_{ij}^k = \tanh((W^k \times x)_{ij} + b_k) \tag{1}$$

As the resolution of FMs gets reduced, the sub-sampling layer is led to spatial invariance to which each pooled FM refers one FM of the prior layer. The sub-sampling function is defined as Eq. (2).

$$\alpha_j = \tanh(\beta \sum_{N \times N} \alpha_i^{n \times n} + b) \tag{2}$$

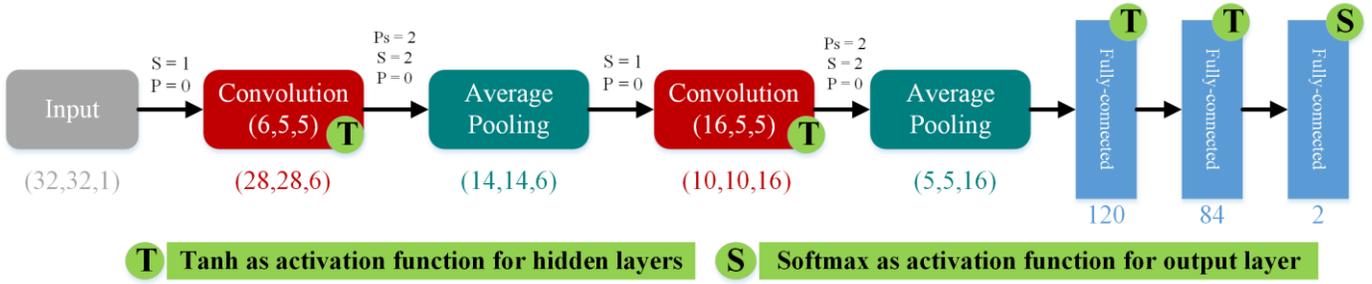

**Fig. 1** The design of LeNet-5 DCNN

**Table 1** The details of the LeNet-5 architecture [52].

| Layers | No. Kernels | Kernel Size | Padding | Stride | Output Features |
|---|---|---|---|---|---|
| Convolution Layer (C1) | 6 | 5×5 | 0 | 1 | (28,28) |
| Average Pooling Layer (S2) | 6 | 2×2 | 0 | 2 | (14,14) |
| Convolution Layer (C3) | 16 | 5×5 | 0 | 1 | (10,10) |
| Average Pooling Layer (S4) | 16 | 2×2 | 0 | 2 | (5,5) |
| Fully Connected Layer (F5) | | | | | 120 |
| Fully Connected Layer (F6) | | | | | 84 |
| Fully Connected Layer (F7) | | | | | 2 |

Where $\alpha_i^{n \times n}$ are the inputs; $\beta$ and $b$ are trainable scalar and bias, respectively. After different convolution and sub-sampling layers, the last layer is a fully connected structure carrying out the classification task. Only a neuron avails for each type of output; therefore, in the case of the COVID-19 dataset, this layer contains two neurons for their types.

## 2.2. Extreme Learning Machine

The ELM is one of the most widely used Single-hidden Layer Neural Network (SLNN) learning algorithms, which its variants are frequently used in sequential learning, batch learning, and incremental learning in consequence of its fast and effective learning speed, appropriate generalization capability, fast convergence rate, and simplicity of implementation [31]. Contrary to canonical learning algorithms, the primary aim of the ELM is to get better generalization performance by achieving both the output weights' smallest norm and the minor training error. As stated in the feedforward neural networks theory of Bartlett [60], the smaller the weights' norm is, the better generalization performance the networks tend to have.

ELM first randomly sets weights and biases of the input layer and then calculates the output layer weights using these random values. This algorithm has a faster learning rate and better performance than the traditional NN algorithms [61]. Fig. 2 indicates a typical SLNN to which $n$ refers to the number of input-layer neurons, $L$ refers to the number of hidden layer neurons, and $m$ refers to the number of output-layer neurons.

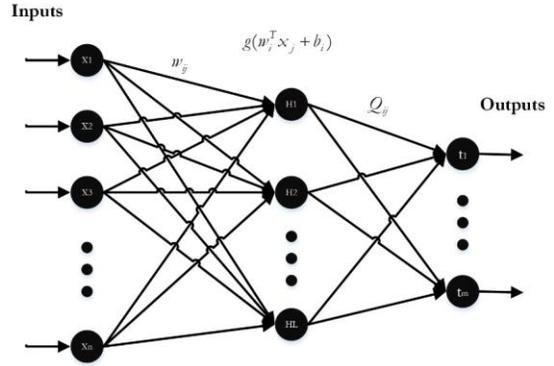

**Fig. 2** Single-hidden layer neural network

As indicated in [33], the activation function can be shown in Eq. (3).

$$\mathbf{Z}j = \sum_{i=1}^{L} Q_i f\left(w_i, b_i, \mathbf{x}_i\right) \tag{3}$$

Where $w_i$ refers to the input weight, $b_i$ refers to the $i$th hidden neuron's bias, $\mathbf{x}_j$ represents the inputs, and $\mathbf{Z}_j$ is the output of the SLNN. Representing matrix of Eq. (3) is shown in Eq. (4).

$$\mathbf{Z}^T = \mathbf{H}Q \tag{4}$$

Where, $Q = [Q_1, Q_2, ..., Q_L]^T$, $\mathbf{Z}^T$ is the transpose of matrix $\mathbf{Z}$, $\mathbf{H}$ is a matrix named as hidden-layer output matrix calculated in Eq. (5).

$$\mathbf{H} = \begin{bmatrix} f(w_1, b_1, \mathbf{x}_1) & f(w_2, b_2, \mathbf{x}_1) & \cdots & f(w_L, b_L, \mathbf{x}_1) \\ \vdots & & \cdots & \vdots \\ f(w_1, b_1, \mathbf{x}_\beta) & f(w_2, b_2, \mathbf{x}_\beta) & \cdots & f(w_L, b_L, \mathbf{x}_\beta) \end{bmatrix}_{\beta \times L} \tag{5}$$

The primary goal of training is to minimize the error or variance of the ELM. Input biases and weights have been stochastically selected. The activation function has to be infinitely differentiable within the conventional ELM, yet in line with this regard, ELM training leads to obtain the output weight (Q) via optimizing the least-squares function indicated in Eq. (6). The corresponding output weights are analytically computed by using the Moore-Penrose generalized in verse as done in ELM (cf. Eq. (7)) instead of any iterative tuning.

$$\min_{Q} \left\| \mathbf{H}Q - \mathbf{Z}^T \right\| \tag{6}$$

$$\hat{Q} = \mathbf{H}^{+}\mathbf{Z}^{T} \tag{7}$$

In this equation, $\mathbf{H}^{+}$ represents the generalized Moore-Penrose inverse of the $\mathbf{H}$ matrix.

Since the ELM's performance is dependent on the number of hidden layer neurons and the number of training epochs, the experiment was conducted with 100 epochs between the number of hidden neurons versus the Root Mean Square Error (RMSE) to specify the best number of hidden neurons. The proposed model's final structure was determined as 120 input neurons, 120 hidden neurons, and output neurons based on the number of classes.

However, due to the random values of input weights and biases, the canonical ELM is not stable enough for real-world engineering problems. Besides, ELM may require a higher number of hidden neurons due to the random determination of the input weights and hidden biases [62,63]. Therefore, optimization algorithms can be employed for tuning input weights and biases to stable the outcomes. Thereby, ChOA is proposed to tune the ELM's input weights and biases in the next section.

### 2.3 The Mathematical Model for ChOA

As proven in [45], ChOA was designed to alleviate the two problems of slow convergence speed and be trapped in local optima compared to other optimization algorithms to solve high-dimensional problems. Considering ELM's parameters tuning dimension, ChOA was utilized to tune the mentioned parameters in this section.

The ChOA is a novel group intelligence-based optimization algorithm inspired by the chimp-hunting mechanism in their communities [45]. Four communal types of chimp are driver, chaser, barrier, and attackers. Although each member of a chimp colony has different capabilities, these differences are necessary for the hunting process. The behavior of the first two charges the driver and the chaser in the hunting group are mathematically defined as follows [46]:

$$\mathbf{d} = \left| \mathbf{c}.\mathbf{x}_{\text{prey}}(t) - \mathbf{m}.\mathbf{x}_{\text{chimp}}(t) \right| \tag{8}$$

$$\mathbf{x}_{\text{chimp}}(t+1) = \mathbf{x}_{\text{prey}}(t) - \mathbf{a}.\mathbf{d} \tag{8}$$

Where $\mathbf{x}_{\text{prey}}$ and $\mathbf{x}_{\text{chimp}}$ represent the position vectors of the prey and the chimp, $t$ refers to the current iteration indicator, and $\mathbf{a}$, $\mathbf{m}$, and $\mathbf{c}$ are the vectors determined by the following equations:

$$\mathbf{a} = 2.\mathbf{f}.\mathbf{r_1} - \mathbf{a} \tag{10}$$

$$\mathbf{c} = 2.\mathbf{r_2} \tag{11}$$

$$\mathbf{m} = \textbf{Chaotic value} \tag{12}$$

Where $f$ is non-linearly reduced over iterations with a range of $[0, 2.5]$, $r1$ and $r2$ are stochastic values between 0 and 1. The chaotic vector $m$ represents the sexual motivations of chimps exploiting different chaotic maps. The stochastic population of the generation of chimps is the first step in the ChOA. The chimps are arranged stochastically into four categories: driver, barrier, attacker, and chaser in the next step. Each group's strategy determines the location of updating method of specified chimps, determining $\mathbf{f}$ vector while all groups attempt to estimate the best position of the prey. The $\mathbf{c}$ and $\mathbf{m}$ vectors are tuned adaptively, and they will improve the local minima avoidance and convergence rate.

Within conventional group intelligence-based meta-heuristic optimization algorithms, different autonomous categories employ different strategies to update $f$, which can be any continuous function as long as it is reduced during iterations [24]. Two kinds of chimps with different autonomous categories named ChOA1 and ChOA2 out of a small number of evaluated methods had the best efficiency in the optimization tasks. The selection for $\mathbf{f}$ in four different groups is presented in Table 2. As it is tabulated, t and T refer to the current and maximum number of iteration, respectively. The attacker chimp leads the exploitation phase, and the hunt is now and then enrolled by the other chimps. Nevertheless, the prey's best location is not accurately determined; the best-obtained solutions are thus used for mathematical modeling of hunting behavior. The first attacker, driver, barrier, and chaser are considered as the best chimps, and the other chimps should update their positions based on these four solutions.

**Table 2** The Mathematical Model for the $\mathbf{f}$ Vector's Dynamic Coefficients [45].

| Category | ChOA1 | ChOA2 |
|---|---|---|
| one | $1.95 - 2 \times t^{1/4}/T^{1/3}$ | $2.5 - (2 \times \log(t)/\log(T))$ |
| Two | $1.85 - 3 \times t^{1/3}/T^{1/4}$ | $(-2 \times t^3/T^3) + 2.5$ |
| Three | $(-3 \times t^3/T^3) + 1.5$ | $0.5 + 2 \times \exp[-(4 \times t/T)^2]$ |
| Four | $(-2 \times t^3/T^3) + 1.5$ | $2.5 + 2 \times (t/T)^2 - 2(2 \times t/T)$ |

Different dynamic strategies are allowed to update $\mathbf{f}$ in autonomous categories to explore search space, having diverse capabilities and homogeneity between local and global search. The adaptively autonomous types provide The following equations represent the position that updates the thumb rule.

$$\mathbf{d}_{\text{Attacker}} = |\mathbf{c_1}\mathbf{x}_{\text{Attacker}} - \mathbf{m_1}\mathbf{x}|, \quad \mathbf{d}_{\text{Barrier}} = |\mathbf{c_2}\mathbf{x}_{\text{Barrier}} - \mathbf{m_2}\mathbf{x}|,$$
$$\mathbf{d}_{\text{Chaser}} = |\mathbf{c_3}\mathbf{x}_{\text{Chaser}} - \mathbf{m_3}\mathbf{x}|, \quad \mathbf{d}_{\text{Driver}} = |\mathbf{c_4}\mathbf{x}_{\text{Driver}} - \mathbf{m_4}\mathbf{x}|. \tag{13}$$

$$\mathbf{x_1} = \mathbf{x}_{\text{Attacker}} - \mathbf{a_1}(\mathbf{d}_{\text{Attacker}}), \quad \mathbf{x_2} = \mathbf{x}_{\text{Barrier}} - \mathbf{a_2}(\mathbf{d}_{\text{Barrier}}),$$
$$\mathbf{x_3} = \mathbf{x}_{\text{Chaser}} - \mathbf{a_3}(\mathbf{d}_{\text{Chaser}}), \quad \mathbf{x_3} = \mathbf{x}_{\text{Driver}} - \mathbf{a_4}(\mathbf{d}_{\text{Driver}}). \tag{14}$$

$$\mathbf{x}(t+1) = \frac{\mathbf{x_1} + \mathbf{x_2} + \mathbf{x_3} + \mathbf{x_4}}{4} \tag{15}$$

Where, $\mathbf{m}$ vector models the chaotic behavior of chimps in the final phase of the hunting to catch more meat which means more social favors such as grooming or sex. Chaotic maps improve convergence rate and avoid local optima entrapping in complex and high-dimensional problems like image processing. Six chaotic maps have been exploited [24], which are deterministic equations with stochastic behaviors as indicated in Table. 2 and Fig. 3. Assume in an algorithm that fifty percent of chimps in the final step of the hunting process will follow their normal behaviors while another fifty percent follow the chaotic strategies to update their successive positions.

**Table 3** Mathematical models of the chaotic maps [45].

| No | Name | Chaotic map | Range |
|----|------|-------------|-------|
| 1 | Chebyshev | $x_{i+1} = \cos(i \times \cos^{-1}(x_i))$ | (-1,1) |
| 2 | Gauss/mouse | $x_{i+1} = \begin{cases} 1 & x_i = 0 \\ \dfrac{1}{\mod(x_i, 1)} & \text{otherwise} \end{cases}$ | (0,1) |
| 3 | Singer | $x_{i+1} = \mu \times (7.86 \times x_i - 23.31 \times x_i{}^2 + 28.75 \times x_i{}^3 - 13.302875 \times x_i{}^4),$ <br> $\mu = 1.07$ | (0,1) |
| 4 | Bernoulli | $x_{i+1} = 2 \times x_i \ (mod\ 1)$ | (0,1) |
| 5 | Sine | $x_{i+1} = \dfrac{a}{4} \times \sin(\pi \times x_i), \quad a = 4$ | (0,1) |
| 6 | Circle | $x_{i+1} = \mod(x_i + b - (\dfrac{a}{2 \times \pi}) \times \sin(2 \times \pi \times x_k), 1), \ \ b = 0.2 \ \text{and} \ a = 0.5$ | (0,1) |

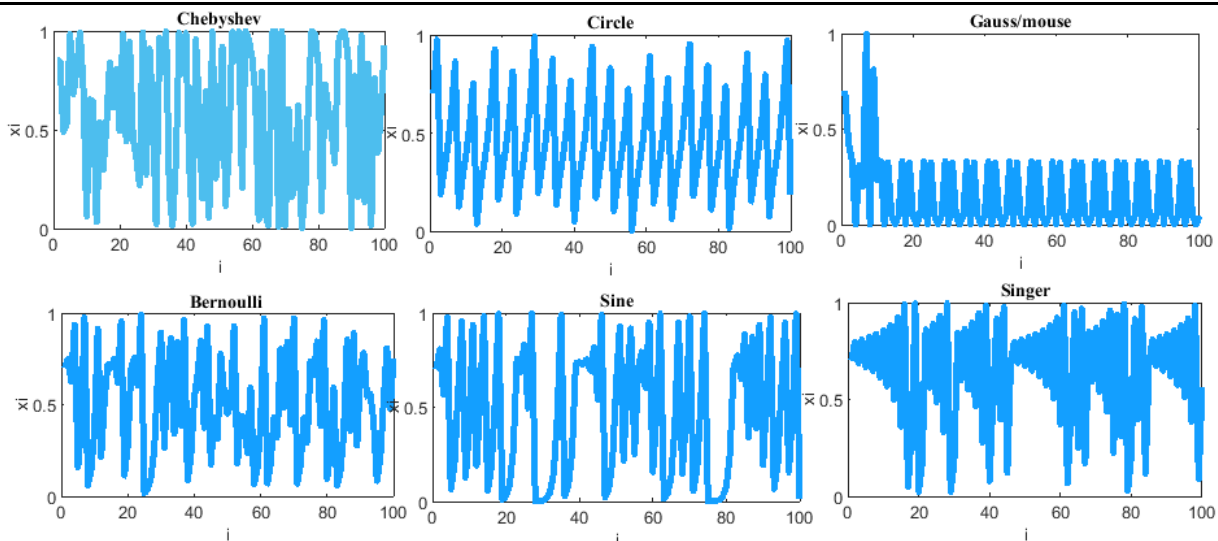

**Fig. 3** The chaotic maps used in ChOA.

The updating model is then mathematically described as in Eq. (16):

$$\mathbf{x}_{chimp}(t+1) = \begin{cases} \mathbf{x}_{prey}(t) - \mathbf{a.d} & if \ \ \mu < 0.5 \\ \text{Chatic\_value} & if \ \ \mu > 0.5 \end{cases} \quad (16)$$

Where $\mu$ is a random number in a range of [0, 1]. Firstly, ChOA is initiated by producing random chimps (candidate solutions). Secondly, all chimps are divided into four mentioned autonomous categories. Thirdly, chimps update their **f** vectors using the assigned categorize strategy. Afterward, the four-categorized chimps evaluate the locations of practicable prey during the iteration. Then, the distances between chimps and the prey will be updated. Moreover, the **c** and the **m** being an adaptive tuning leads to a local optima avoidance and a faster convergence rate simultaneously. Finally, chaotic maps lead to accelerating the convergence rate while avoiding local minima.

### 2.4 COVID-19 Dataset

In this study, two datasets were utilized to evaluate the performance of the designed model. The first one is the *COVID-X-ray-5k* dataset comprises 2084 training samples and 3100 test images [64]. In this dataset, considering radiologist advice, only anterior and posterior COVID-19 X-ray images are used since the lateral photos are not applicable for purposefully detecting. Expertise radiologists evaluated those images and eliminated those ones not having clear pieces of evidence for COVID-19. The *COVID-X-ray-5k* dataset includes 224316 chest X-ray images from 65240 patients. 2000 and 3000 non-COVID images were chosen from this dataset for the training and the testing sets, respectively. In this way, 19 images out of 203 images were removed, and 184 images remained, indicating clear pieces of evidence of COVID19. A group of a more clearly labeled dataset was introduced in this method. 100 images out of 184 photos are considered for the test set, and 84 images are intended for the training set. For increasing the number of positive cases to 420, data augmentation is applied. Since the number of normal cases was small in the covid-chestxray-dataset [64,65], the supplementary ChexPert dataset [66]was employed. The second dataset is extracted from the *COVIDetectioNet* study, produced based on publicly available X-ray image datasets [67]. Three X-ray image datasets obtained from the Kaggle and Github databases were utilized to generate this hybrid dataset. The COVIDetectioNet dataset includes chest X-ray images of COVID-19, Normal, and Pneumonia cases, with 219, 1583, and 4290 samples, respectively. Table 4 represents the detailed information on the utilized datasets and their sources. Fig. 4 indicates some stochastic sample cases from utilized datasets, including normal, Pneumonia, and COVID-19 samples. The final number of images related to different classes is reported in Table 5.

**Table 4** The detailed information on the utilized datasets and their sources

| Dataset | Normal | COVID-19 | Pneumonia | Total |
|---|---|---|---|---|
| **Source datasets** | | | | |
| covid-chestxray-dataset[1] | – | 76 | 17 | 93 |
| COVID-19 Radiography Database[2] | 1341 | 219 | 1345 | 2905 |
| Chest X-Ray Images (Pneumonia)[3] | 1583 | – | 4273 | 5856 |
| **Used datasets** | | | | |
| *COVIDetectioNet* | 1583 | 219 | 4290 | 6092 |
| *COVID-X-ray-5k* | 5000 | 520 | - | 5520 |

**Table 5** The Categories of Images per Class in the utilized Datasets.

| Category | COVID19 | Normal |
|---|---|---|
| ***COVID-X-ray-5k*** | | |
| Training Set | 420 (84 before augmentation) | 2000 |
| Test Set | 100 | 3000 |
| ***COVIDetectioNet*** | | |
| Training Set | 150 | 2873 |
| Test Set | 69 | 3000 |

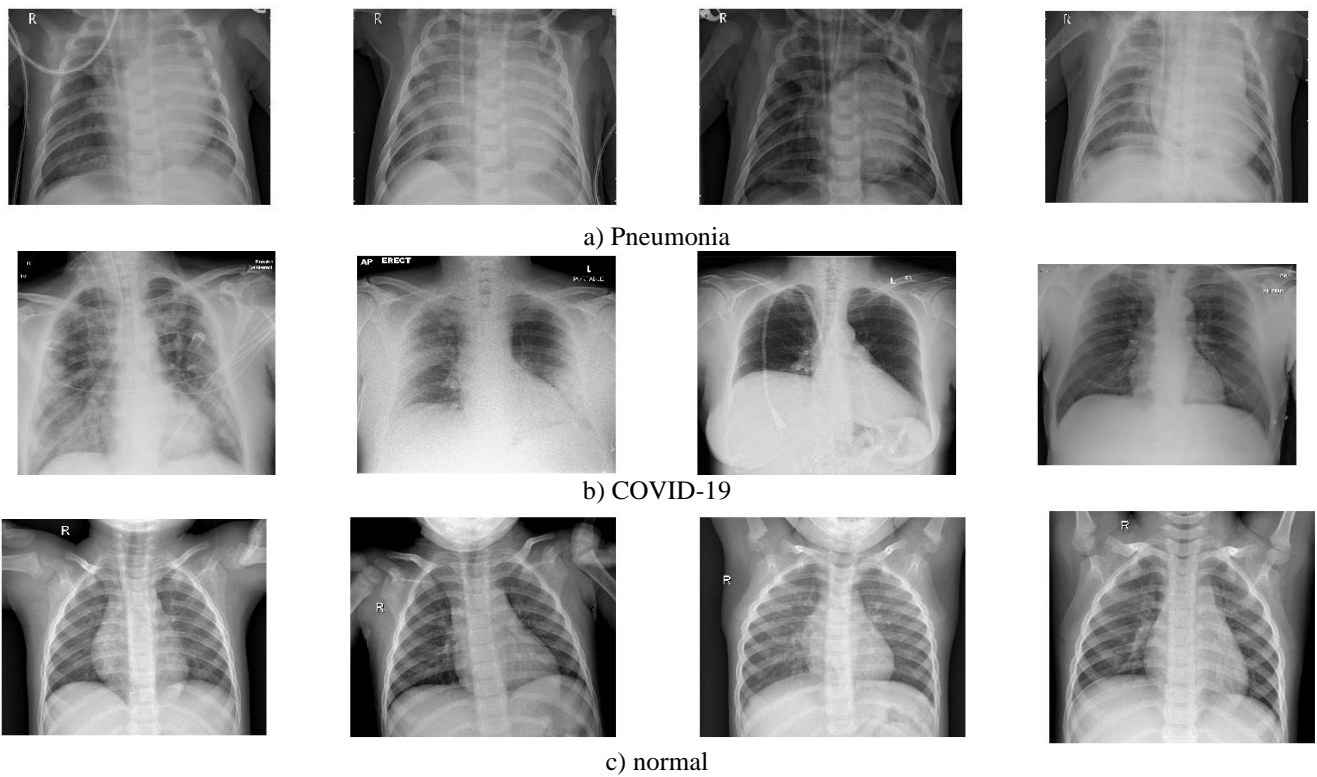

a) Pneumonia

b) COVID-19

c) normal

**Fig. 4** Some samples of a) Pneumonia, b) COVID-19, and c) normal cases from the utilized datasets.

### 3. Methodology

This paper uses the LetNet-5 structure to detect COVID-19 positive cases. It consists of three convolutional layers and two pooling layers followed by a Fully-Connected (FC) layer using Gradient Descent-based Back Propagation (GDBP) algorithm for learning [68]. Regarding the aforementioned GDBP deficiencies, we propose to use a single-layer ELM instead of FC layers to classify the extracted features, as shown in Fig. 5.

[1] Source: https://github.com/tawsifur/COVID-19-Chest-X-ray-Detection
[2] Source: https://www.kaggle.com/tawsifurrahman/covid19-radiography-database
[3] Source: https://www.kaggle.com/paultimothymooney/chest-xray-pneumonia

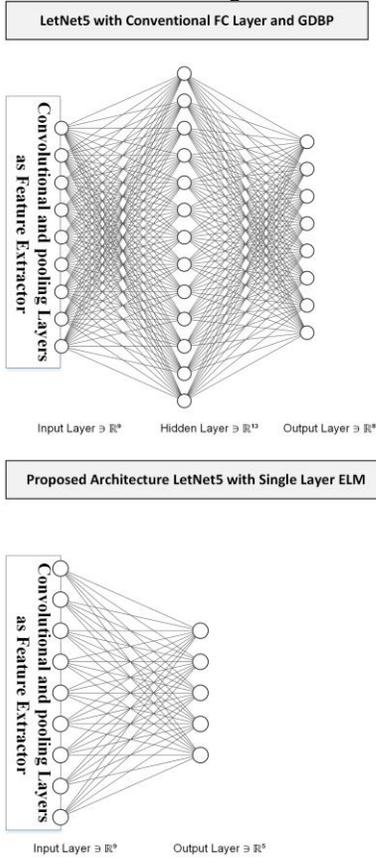

**Fig. 5** the Conventional vs. Proposed Design

The convolutional layers' weights are pre-trained on a large dataset as a complete LetNet-5 with a standard GDBP learning algorithm. After the pre-training phase, the FC layers are removed and remained layers are frizzed to exploit as a feature extractor. The ELM network's input values will be provided by the features generated by the stub- CNN. The ELM has 120 hidden-layer neurons and two output neurons in the proposed structure. It is to notify that the sigmoid function is used as an activation function.

### 3.1 **Stabilizing ELM Using ChOA**

Despite the reduction of training time in ELMs compared to the standard FC layer, ELMs are not stable and reliable in real-world engineering problems due to the random determination of the input layer's weights and biases. As proven in [45], ChOA was designed to alleviate the two problems of slow convergence speed and get trapped in local optima compared to other optimization algorithms in solving high-dimensional problems. Considering ELM's parameters tuning dimension, we apply the ChOA to adapt the input layer weights and biases of ELM to increase the network's stability and reliability (ChOA-ELM) while keeping the real-time operation.

Generally speaking, there are two main issues in adapting (tuning) a deep network using a meta-heuristic optimization algorithm. First, the structure's parameters have been clearly represented by the searching agents (candid solution) of the meta-heuristic algorithm; second, the fitness function must be defined based on the problem's interest.

The presentation of network parameters is a distinctive phase in tuning a Deep Convolutional ELM using ChOA (DCELM-ChOA) algorithm. Thereby, ELM's input layer's weights and biases should be clearly determined to make the best diagnostic accuracy. By and large, ChOA optimizes ELM's input layer's weights and biases, which are used to calculate the loss function as a fitness function. In fact, the values of weight and bias are used as searching agents (Chimps) in the ChOA.

Three schemes are generally used to present weights and biases of a DCELM as candid solutions for the meta-heuristic algorithm: vector-based, matrix-based, and binary state [69-71]. Since the ChOA needs parameters in a vector-based model, the candid solution is shown as Eq. 10.

$$\mathbf{Chimps} = [W_{11}, W_{12}, ...., W_{nL}, b_1, ...., b_L] \qquad (17)$$

Where $n$ is the number of the input nodes, $W_{ij}$ indicates the connection weight between the $i_{th}$ feature node, and $j_{th}$ refers to the input neuron of ELM, $b_j$ is the bias of the $j_{th}$ input neuron. As previously stated, the proposed design is a simple LeNet-5 structure [52]. In this section two structures named as $in\_6c\_2p\_12c\_2p$ and $in\_8c\_2p\_16c\_2p$ are used whereas $c$ and $p$ are convolution and pooling layers respectively. The kernel size of all convolution layers is 5x5, and the scale of pooling is down-sampled by a factor of 2.

### 3.2 **Loss Function**

In the proposed meta-heuristic method, the ChOA algorithm trains DCELM to obtain the best accuracy and minimize evaluated classification errors. This aim can be computed by the loss function of the metaheuristic searching agent or the Mean Square Error (MSE) of the classification procedure. Albeit, the loss function used in this method is as follows [72]:

$$E = \sqrt{\frac{\sum_{j=1}^{N} \mathbf{P} \sum_{i=1}^{k} \mathbf{Q}_{i.} f(\mathbf{w}_i.\mathbf{x}_j + b_i) - \mathbf{t}_j \, \mathbf{P}_2^2}{m \times N}} \qquad (18)$$

Where $n$ is the number of the input nodes, $W_{ij}$ indicates the connection weight between the $i_{th}$ feature node and $j_{th}$ refers to the input neuron of ELM, $b_j$ is the bias of the $j_{th}$ input neuron, $\mathbf{x}_j$ represents the inputs, $Q$ denotes the output weight, and $N$ refers to the number of training samples. The proposed ChOA algorithm uses two termination criteria, including reaching maximum iteration or pre-defined loss function. Consequently, the general block diagram and the pseudo-code of DCELM-ChOA are shown in Fig. 6 and Fig. 7, respectively.

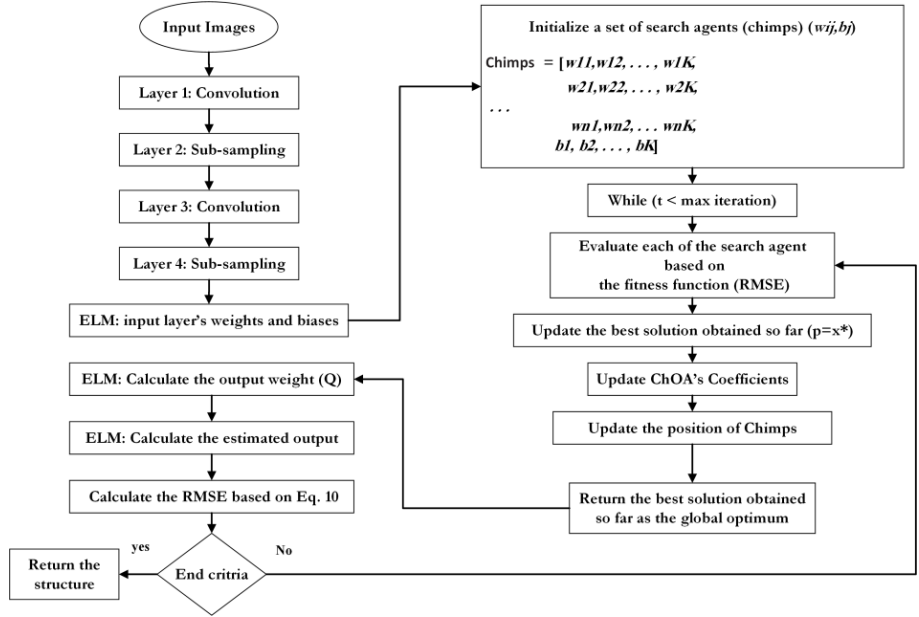

**Fig. 6** A general block diagram of the CELM-ChOA model.

| **Pseudo-Code DCELM -ChOA** |
| --- |

**Result**: computation *time, and classification accuracy,*
 Initialization of the DCELM structure
Calculation process*: loss function f(x), connection weights (w), and biases (b̂)*
 **Initialize** *the population of chimps (candid solution) $x_i$ (i=1,2, ...,n)*
**Initialize f, m, a** *and* **c**
Calculate the chimp's position
Divide chimps randomly into autonomous categories
***Until*** *stopping condition is satisfied*
 Calculate the loss function for each chimp
  $\mathbf{x}_{\text{Attacker}}$ = *the best chimp*
  $\mathbf{x}_{\text{Chaser}}$ = *the second-best chimp*
  $\mathbf{x}_{\text{Barrier}}$ = *the third-best chimp*
  $\mathbf{x}_{\text{Driver}}$ = *the fourth-best chimp*
 **while** *(t < maximum number of iterations)*
  **for** *each chimp:*
   *Extract the chimp's category number*
   *Use its category equation to update* **f, m,** *and* **c**
   *Use* **f, m,** *and* **c** *to calculate* **a** *and then* **d**
  **end for**
   **for** *each chimp*
    **if** *( $\mu$ <0.5)*
     **if** *(|a| < 1)*
*Update the chimp's position by Eq. (2)*
     **else if** *(|a|>1) Select a chaotic candid solution*
     **end if**
    **else if** *( $\mu$ >0.5)*
     *Update the chimp's position using Eq.(9)*
    **end if**
  **end for**
  *Update* **f, m, a** *and* **c**
  *Update* $\mathbf{x}_{\text{Attacker}}$, $\mathbf{x}_{\text{Driver}}$, $\mathbf{x}_{\text{Barrier}}$, $\mathbf{x}_{\text{Chaser}}$
  *t=t+1*
 **end while**
**return** $\mathbf{x}_{\text{Attacker}}$

**Fig. 7** The Pseudo-code for DCELM-ChOA model.

## 4. Simulation Results and Discussion

The hybrid method's initial target is to enhance the diagnosis rate of classic DCNN by using the ELM and the ChOA learning algorithm. In the DCELM-ChOA simulation, the population and maximum iteration equal to 50 and 10, respectively. The parameter of DCNN, i.e., the learning rate $\alpha$ and the batch size equal to 0.0001 and 20 accordingly. Additionally, the number of epochs is considered between 1 and 10 for every evaluation. We down-sample all input images to 31×31 before applying them in DCNNs. The evaluation was run in the MATLAB-R2019a on a PC with Intel Core i7-4500u processor 16 GB RAM in Windows 10 with five individual runtimes. The performance of DCELM-ChOA is compared with DCELM [73], DCELM-GA [27], DCELM-CS [74], and DCELM-WOA [75] on the utilized datasets. The parameters of the GA, the CS, the DA, and the WOA are shown in Table 6.

**Table 6** The Parameters of Benchmark Algorithms

| Algorithm | Parameters | Values |
| --- | --- | --- |
| GA | Cross-over Probability | 0.7 |
| | Mutation Probability | 0.1 |
| | Population Size | 50 |
| CS | Discovery Rate of Alien Eggs | 0.25 |
| | Population Size | 50 |
| WOA | *a* Linearly Decreased from 2 to 0 | |
| | Population Size | 50 |
| ChOA | **f** | Table 1 |
| | **m** | Chaotic |
| | $\mathbf{r_1}$, $\mathbf{r_2}$ | Random |
| | Population Size | 50 |

### 4.1. Evaluation Metrics

Different metrics can be remarkably used to measure the classification model's efficiency, such as sensitivity, classification accuracy, specificity, precision, Gmean,

Norm, and F1-score. Since the datasets are significantly imbalanced (169 COVID19 images, 6000 NonCOVID images), we use specificity (true negative rate) and sensitivity (true positive rate) to report the performance of designed models as following Eqs correctly [76]:

$$Sensitivity\ (TPR) = \frac{TP}{P} = \frac{TP}{TP + FN} \qquad (19)$$

$$Specificity\ (TNR) = \frac{TN}{N} = \frac{TN}{TN + FP} \qquad (20)$$

Where, TP represents the number of true positive cases, FN represents the number of false-negative cases, TN represents the number of true negative cases, and FP points to the number of false-positive cases.

### 4.2. The Analysis of Chaotic maps' effects

This subsection evaluates the sensitivity analysis of chaotic maps employed in the ChOA on the overall performance. Considering the references [63,77], experiments were conducted using six chaotic maps (i.e., Chebyshev, Gauss/mouse, Singer, Bernoulli, Sine, Circle) defined in Table 2. The designed model is trained for each chaotic map. The calculated classification accuracy for chaotic maps is represented in Fig. 8. In which, Fig. 8a indicates the results for the *COVID-X-ray-5k* dataset, and Fig. 8b shows the results for the *COVIDetectioNet* dataset. As shown in this figure, the best performance is obtained for Gauss/mouse map. Therefore, this chaotic map is chosen for the following comparison.

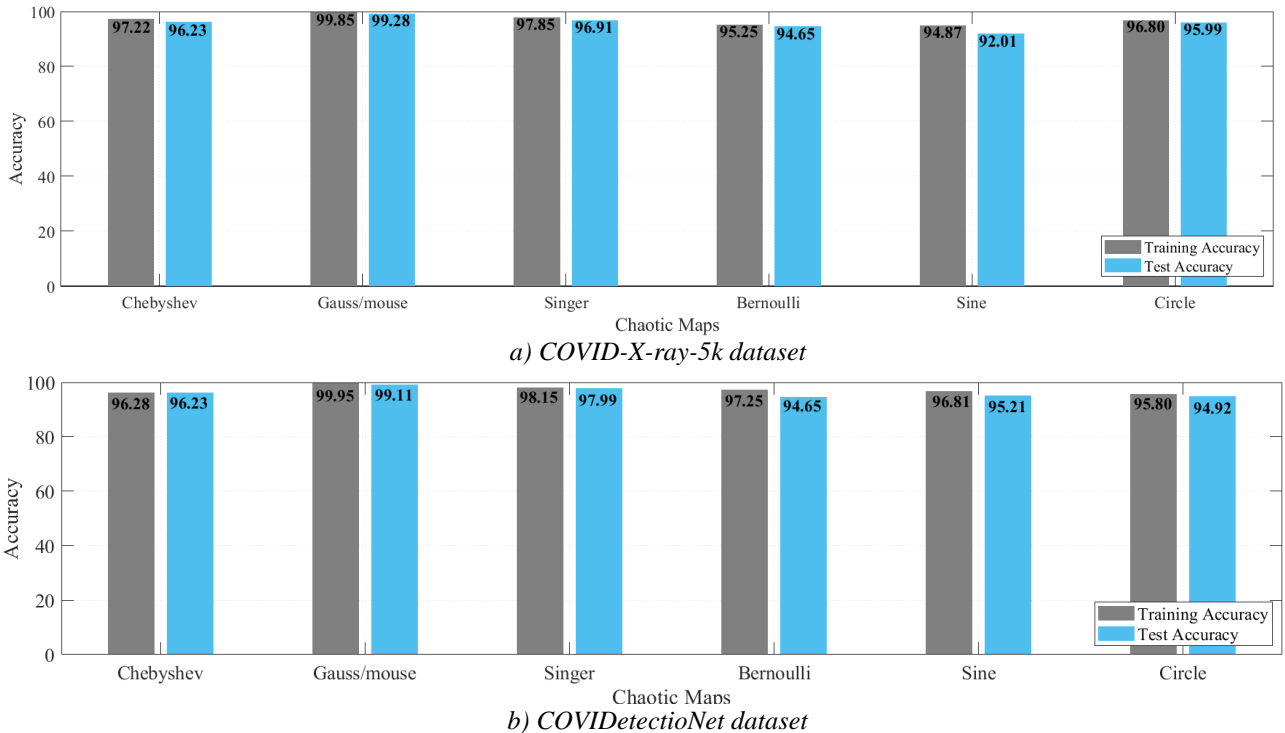

a) COVID-X-ray-5k dataset

b) COVIDetectioNet dataset

Fig. 8 The classification accuracy for different chaotic maps.

### 4.3. Structure Expected Probability Grades

As for the importance of time complexity we use two simple LetNet-5 convolutional structures, i. e., in_6c_2p_12c_2p and in_8c_2p_16c_2p. The probability of each image is predicted by these structures, indicating the possibility of the image being identified as COVID-19. As comparing this similarity with a threshold, we can extract a binary label that indicates if the specified image is COVID-19 or not. A perfect structure must identify the similarity of all COVID-19 cases close to one and Non-Covid cases close to zero.

Fig. 9 and Fig. 10 display Expected Probability Grades (EPG) distribution for the images in the *COVID-X-ray-5k and COVIDetectioNet* test datasets. Since the Non-Covid category comprises general cases and other types of infection, the EPG distribution is presented for three categories, i.e., COVID-19, Non-COVID, and Pneumonia

Cases. As shown in Fig. 9 and Fig. 10, the Pneumonia Cases have slightly larger grades than the Non-COVID cases. That the Pneumonia images are more complex to be recognized from COVID-19 than Non-COVID general cases is logical. Positive COVID-19 cases are expected to have much higher probabilities than the Non-COVID cases, certainly stimulating, as it indicates that the structure is learning to recognize COVID-19 from Non-COVID samples. The confusion matrices for these two structures on *COVID-Xray-5k* and *COVIDetectioNet* are shown in Figs. 11 and 12.

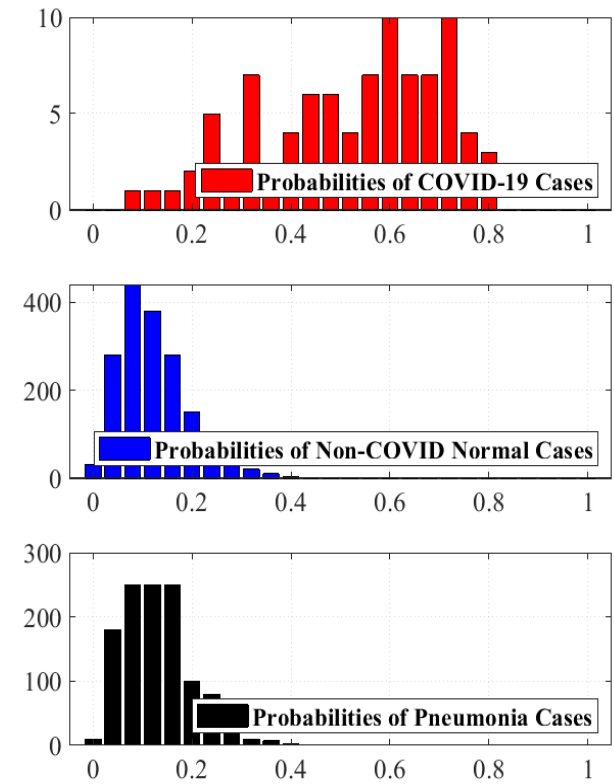

*a*) in_6c_2p_12c_2p Structure

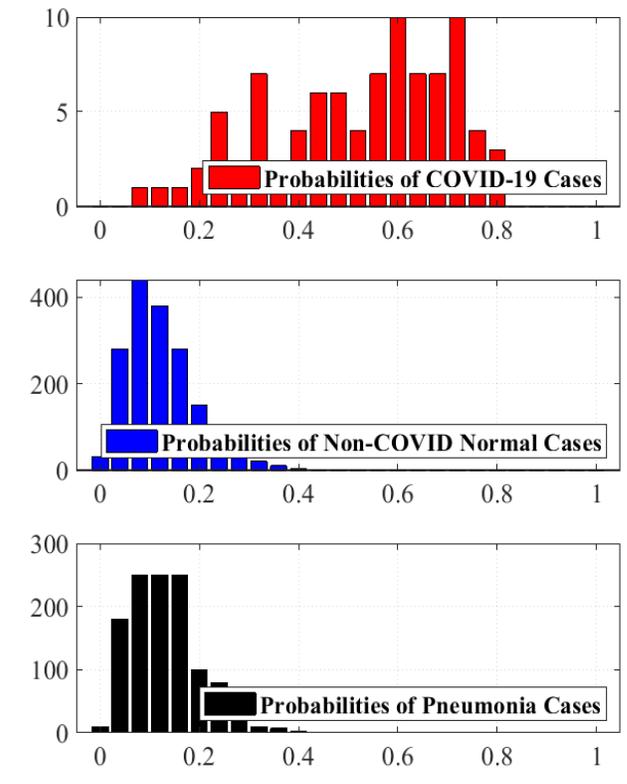

*a*) in_6c_2p_12c_2p Structure

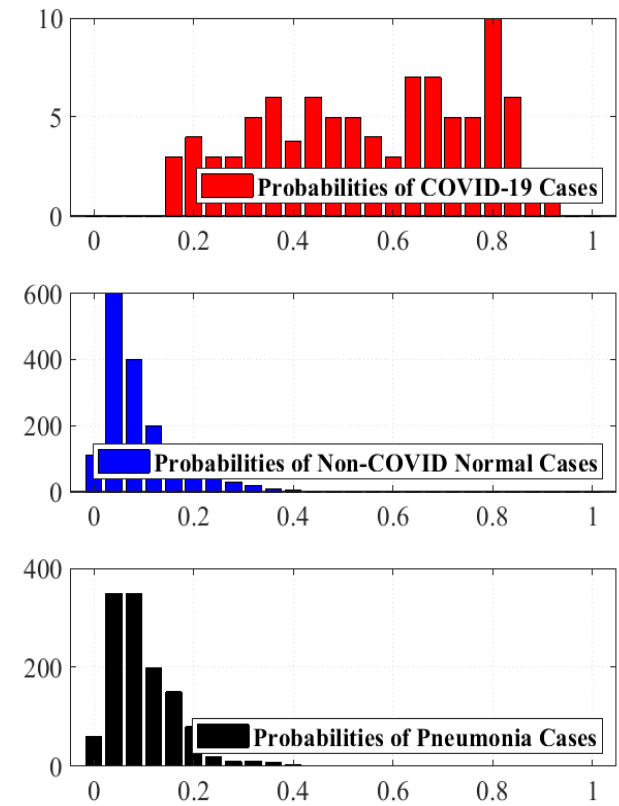

*b*) in_8c_2p_16c_2p Structure

**Fig. 9** the EPG for *COVID-Xray-5k dataset*

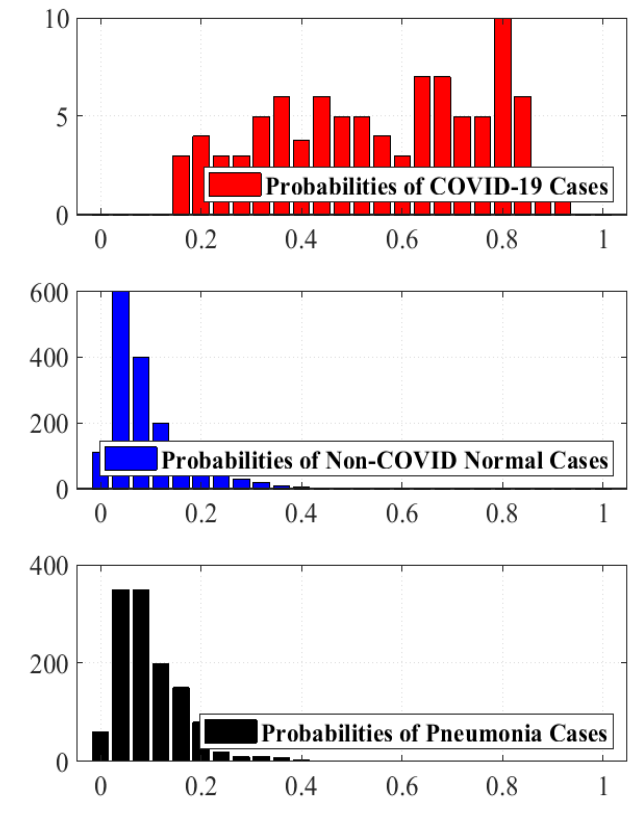

*b*) in_8c_2p_16c_2p Structure

**Fig. 10** the EPG for *COVIDetectioNet dataset*

## in_6c_2p_12c_2p structure

a) in_6c_2p_12c_2p structure

## in_8c_2p_16c_2p structure

b) in_8c_2p_16c_2p structure

**Fig. 11** The Confusion Matrix for *COVID-Xray-5k dataset*

## in_6c_2p_12c_2p structure

a) in_6c_2p_12c_2p structure

## in_8c_2p_16c_2p structure

b) in_8c_2p_16c_2p structure

**Fig. 12** The Confusion Matrix for *COVIDetectioNet dataset*

Considering the calculated results, we opt for the in_8c_2p_16c_2p structure as a benchmark structure named as conventional DCNN.

### 4.4. The Comparison of Specificity and Sensitivity

Each structure of the EPG indicates the possibility of the image being COVID-19. These EPGs can be compared with a cut-off threshold to deduce if the image is a positive COVID-19 case or not. Researchers use calculated labels to evaluate the specificity and sensitivity of each detector. Different specificity and sensitivity rates can be calculated based on the value of the cut-off threshold. The specificity and the sensitivity rates based on conventional DCNN, DCELM, DCELM-GA, DCELM-CS, DCELM-WOA, and DCELM-ChOA models are represented in Table 7 and Table 8 for *COVID-Xray-5k* and *COVIDetectioNet* datasets, respectively.

**Table 7** The Specificity and Sensitivity Rates of Benchmark Models for *COVID-Xray-5k* dataset.

| Model | Threshold | Sensitivity (%) | Specificity (%) |
|---|---|---|---|
| **DCNN** | 0.1 | 98 | 84.47 |
| | 0.2 | 95 | 85.73 |
| | 0.3 | 90 | 87.42 |
| | 0.4 | 84 | 90.82 |
| **DCELM** | 0.1 | 98 | 83.37 |
| | 0.2 | 94 | 86.21 |
| | 0.3 | 89 | 88.12 |
| | 0.4 | 83 | 89.52 |
| **DCELM-GA** | 0.1 | 98 | 92.26 |
| | 0.2 | 97 | 93.85 |
| | 0.3 | 92 | 94.85 |
| | 0.4 | 89 | 96.85 |
| **DCELM-CS** | 0.1 | 99 | 89.91 |
| | 0.2 | 97 | 92.85 |
| | 0.3 | 95 | 96.33 |
| | 0.4 | 91 | 97.33 |
| **DCELM-WOA** | 0.1 | 99 | 85.12 |
| | 0.2 | 96 | 92.98 |
| | 0.3 | 91 | 96.60 |
| | 0.4 | 80 | 97.90 |
| **DCELM-ChOA** | 0.1 | 100 | 84.34 |
| | 0.2 | 98 | 93.32 |
| | 0.3 | 97 | 95.33 |
| | 0.4 | 92 | 98.66 |

**Table 8** The Specificity and Sensitivity Rates of Benchmark Models for *COVIDetectioNet* dataset.

| Model | Threshold | Sensitivity (%) | Specificity (%) |
|---|---|---|---|
| **DCNN** | 0.1 | 97 | 84.32 |
| | 0.2 | 96 | 86.11 |
| | 0.3 | 91 | 87.93 |
| | 0.4 | 84 | 90.55 |
| **DCELM** | 0.1 | 98 | 82.92 |
| | 0.2 | 95 | 87.01 |
| | 0.3 | 90 | 87.98 |
| | 0.4 | 83 | 89.44 |
| **DCELM-** | 0.1 | 98 | 91.95 |
| | 0.2 | 97 | 94.11 |
| | 0.3 | 93 | 95.07 |

| GA | 0.4 | 90 | 97.01 |
|---|---|---|---|
| | 0.1 | 99 | 90.01 |
| | 0.2 | 98 | 93.05 |
| DCELM-CS | 0.3 | 96 | 96.91 |
| | 0.4 | 92 | 97.82 |
| | 0.1 | 98 | 84.44 |
| | 0.2 | 97 | 93.05 |
| DCELM-WOA | 0.3 | 92 | 96.82 |
| | 0.4 | 81 | 98.17 |
| | 0.1 | 100 | 85.14 |
| | 0.2 | 99 | 94.01 |
| DCELM-ChOA | 0.3 | 97 | 95.99 |
| | 0.4 | 93 | 98.89 |

The data presented in Table 7 and 8 show that all benchmark networks obtain much favorable outcomes, and the best performing structure (DCELM-ChOA) achieves a sensitivity rate of 100% and a specificity rate of 98.66% for *COVID-Xray-5k* datasets and sensitivity rate of 100% and a specificity rate of 98.89% for *COVIDetectioNet* datasets. DCELM-ChOA and DCELM-WOA become slightly better in efficiency than other benchmark structures.

### 4.5. The Reliability Analysis of Imbalance Dataset

Considering the limitation of the number of approved labeled positive COVID-19 cases, the researchers have only 100 positive COVID-19 cases put in the *COVID-Xray-5k* dataset; that is why sensitivity and specificity rates, which are reported in Table 5, might not be completely reliable. Theoretically, more numbers of positive COVID-19 cases are needed to conduct a more reliable evaluation of sensitivity rates. Albeit, 95% confidence interval of the obtained specificity and sensitivity rates can be evaluated to test what is the feasible interval of calculated values for the current number of test cases in each category. The confidence of interval for the accuracy rate can be calculated as in Eq. 21 [78,79].

$$r = p \sqrt{\frac{\text{Accuracy.Rate}(1\text{-Accuracy.Rate})}{N}} \qquad (21)$$

Where, *p* refers to the significance level of the confidence interval, i.e., Standard Deviation (SD) of the Gaussian distribution, *N* refers to the number of cases for each class, Accuracy.Rate refers to the evaluated accuracy, sensitivity, and specificity in this example. The 95% used confidence interval is to lead the corresponding value of 1.96 top. Regarding the fact that a sensitive network is essential for the COVID-19 detection problem, the particular threshold levels are selected corresponding to a sensitivity rate of 98% for each benchmark network, and their specificity rates are examined afterward. The comparison of the six model's performance is presented in Table 9 and Table 10 for *COVID-Xray-5k* and *COVIDetectioNet* datasets, respectively. The data presented in these tables show that the specificity rates' confidence interval is about 1 %. Comparatively, it equals around 2.8 % for sensitivity since there are 3000 images for the normal classes (Non-COVID and Pneumonia).

**Table 9** The Reliability Analysis of Sensitivity and Specificity for *COVID-Xray-5k* dataset.

| Model | Sensitivity (%) | Specificity (%) |
|---|---|---|
| **DCNN** | $98 \pm 2.8$ | $84.47 \pm 1.31$ |
| **DCELM** | $98 \pm 2.8$ | $83.37 \pm 1.32$ |
| **DCELM-GA** | $98 \pm 2.8$ | $92.26 \pm 0.90$ |
| **DCELM-CS** | $98 \pm 2.8$ | $91.85 \pm 0.91$ |
| **DCELM-WOA** | $98 \pm 2.8$ | $91.33 \pm 0.91$ |
| **DCELM-ChOA** | $98 \pm 2.8$ | $93.32 \pm 0.89$ |

**Table 10** The Reliability Analysis of Sensitivity and Specificity for *COVIDetectioNet* dataset.

| Model | Sensitivity (%) | Specificity (%) |
|---|---|---|
| **DCNN** | $98 \pm 2.8$ | $85.11 \pm 1.28$ |
| **DCELM** | $98 \pm 2.8$ | $83.22 \pm 1.29$ |
| **DCELM-GA** | $98 \pm 2.8$ | $91.92 \pm 0.89$ |
| **DCELM-CS** | $98 \pm 2.8$ | $92.03 \pm 0.92$ |
| **DCELM-WOA** | $98 \pm 2.8$ | $91.15 \pm 0.89$ |
| **DCELM-ChOA** | $98 \pm 2.8$ | $94.02 \pm 0.88$ |

Comparing different structures solely based on their specificity and sensitivity rates does not make enough sense of the detector's performance because different threshold levels cause different specificity and sensitivity rates. The precision-recall curve is a good presentation that can be used to evaluate comparison between these networks for all feasible cut-off threshold levels comprehensively. This presentation shows the precision rate as a function of the recall rate. Precision is then defined as the TPR divided by the TP (i.e., Eq. 19), and the recall has the same definition as TNR (i.e., Eq. 20). Figs. 13 and 14 show the precision-recall plot of these six benchmark models. The Receiver Operating Characteristic (ROC) plot is another appropriate tool showing the TPR as a function of FPR. Therefore, these figures show the ROC curve of these six benchmark structures as well. The ROC curves show that DCELM-ChOA significantly outperforms other DCELM-based networks and yet as well as conventional DCNN on the test dataset. It comes to notify that the Area Under Curve (AUC) of ROC curves might not rightly indicate the model's efficiency since it can be very high for widely imbalanced test sets, including *COVID-Xray-5k* and *COVIDetectioNet* datasets.

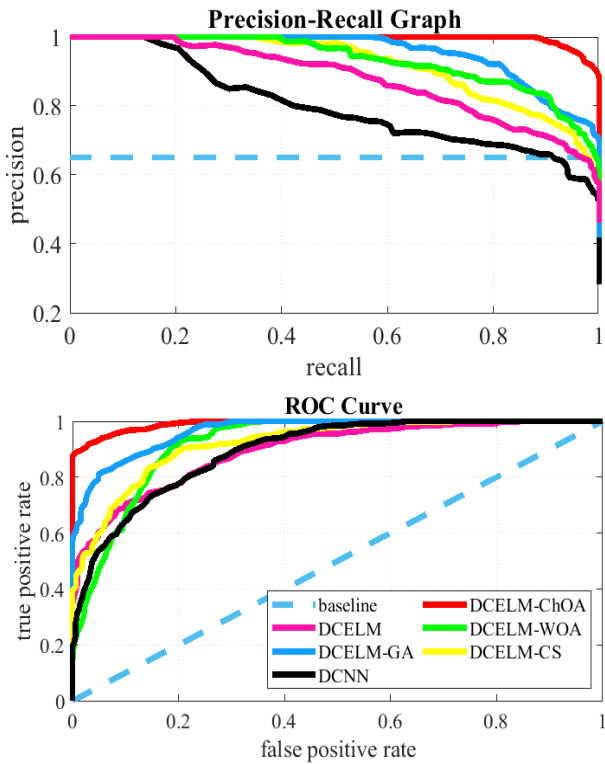

**Fig. 13** The ROC Curves and Precision-recall Curves for *COVID-Xray-5k* dataset

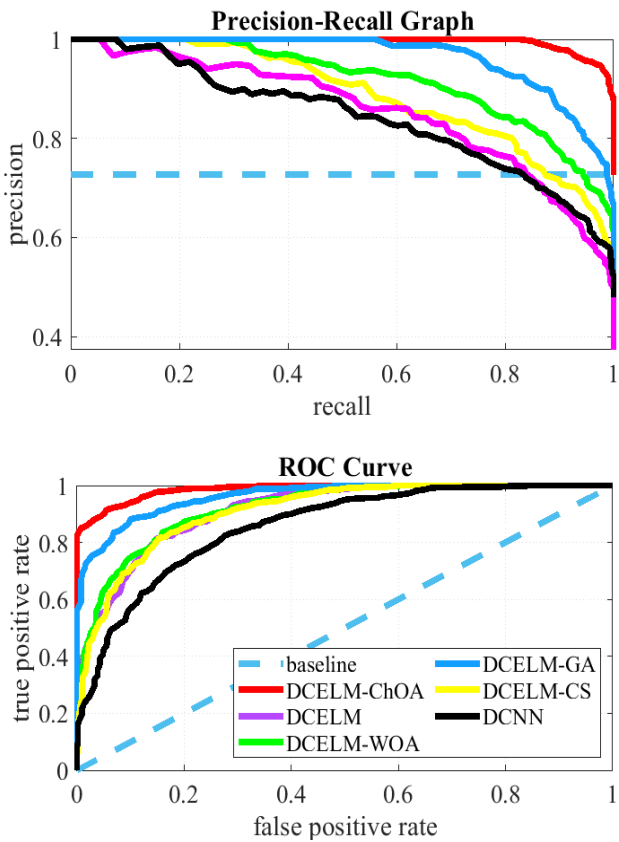

**Fig. 14** The ROC Curves and Precision-recall Curves for *COVIDetectioNet* dataset

As figures are to show the results, the DCELM-ChOA detector presents significant COVID-19 detection as it is compared with other benchmark models. The proposed approach outperforms other comparative benchmarks with 98.25% and 99.11% as ultimate accuracy on the *COVID-Xray-5k* and *COVIDetectioNet* datasets, respectively, and it led relative error to reduce as the amount of 1.75% and 1.01% as compared to a convolutional CNN.

Generally, the precision-recall plot shows the trade-off between recall and precision for different threshold levels. A high area under the precision-recall curve represents high precision and recall. High precision represents a low false-positive rate and high-recall represents a low false-negative rate. As it can be observed from the curves in Figs. 13 and 14, the DCELM-ChOA has a higher area under the precision-recall curves; it thus means a lower false positive and false negative rate than other benchmark detectors. The results of the simulation indicate that DCELM-ChOA represents the best accuracy for all epochs.

As shown from the ROC and precision-recall curves, the AUC of DCELM is reduced compared to the conventional DCNN.

This reduction means that the performance of DCNN slightly decreases when the ELM is replaced with the fully connected layer since the merits of supervised learning are ignored. Nevertheless, it is stated that the other evolutionary DCNNs have better performance than standard DCNN. In reality, the advantage of the stochastic supervised nature of the evolutionary learning algorithm and unsupervised nature of the ELM is taken. Consequently, the detector's performance is improved as advantages of these hybrid-supervised and -unsupervised learning algorithms are compounded.

### 4.6. The Analysis of Time Complexity

Measuring the time complexity is necessary for analyzing a real-time detector. Regarding the benchmark networks, researchers run the designed COVID-19 detector using NVidia Tesla K20 as the GPU and an Intel Core i7-4500u processor as the CPU. The testing time is the time demanded to process the whole test set of 3100 images. We use a non-parametric statistical procedure, Wilcoxon's rank-sum test [80,81], at 5% significance level to test whether the outcomes of DCELM-ChOA differ from other benchmark networks in a statistically significant way or not. The p-values are tabulated in Table 11 as well. In this Table, N/A represents "Not Applicable," i.e., in Wilcoxon's rank-sum test, the corresponding network cannot be compared with itself. As notified, p-values less than 0.05 are referred to as strong evidence against the null hypothesis. It is also necessary to assume that p-values greater than 0.05 are underlined. It should be noted that the results in Table 11 are the average of the results of the two data sets used.

**Table 11** The Comparison of Test and Training Time of Benchmark Network Run on GPU and CPU.

| Model | CPU vs. GPU | Training time | Testing time | P-value |
|---|---|---|---|---|
| DCNN | GPU | 11 min, 11 sec | 3152 ms | 2.11E-07 |
|  | CPU | 6 h, 32 min, 7 sec | 4 min, 29 sec | 1.41E-03 |
| DCELM | GPU | **1162 ms** | **2929 ms** | **N/A** |
|  | CPU | **1 min, 14 sec** | **4 min, 03 sec** | **N/A** |
| DCELM-GA | GPU | 3632.7 ms | 3107 ms | 1.21E-05 |
|  | CPU | 4 min, 27.5 sec | 4 min, 23 sec | 1.11E-05 |
| DCELM-CS | GPU | 2582.3 ms | 3102 ms | 1.58E-04 |
|  | CPU | 3 min, 9.6 sec | 4 min, 28 sec | 1.29E-04 |
| DCELM-WOA | GPU | 1298.9 ms | 3014 ms | **0.505** |
|  | CPU | 2 min, 5 sec | 4 min, 19sec | 1.37E-08 |
| DCELM-ChOA | GPU | 1233 ms | 2932 ms | **0.538** |
|  | CPU | 1 min, 69 sec | 4 min, 19sec | **0.513** |

From another point of view and based on the result of Table 11 taken, it is crystal clear that the training and the testing time of DCELMs are remarkably lower than the classic DCNN. It is also noteworthy that in GPU accelerated training, the proposed approach 538 times faster than the current DCNN. Perpending the number of testing and training images in Table 3 and also revolving of time of entire test and training processing in Table 11 that the DCELMs requires less than one millisecond per image for both training and testing can easily be ramified and thus makes DCELMs perform in real-time in both phases. Since more than 90% of the processing time is related to feature extraction, using other deep learning models can reduce processing time even further.

### 4.7 Identifying the Region of Interest

From the data science experts' perspective, the best result can be shown under the confusion matrix, overall accuracy, precision, recall, ROC curve, etc. [82]. However, these optimal results might not be sufficient for medical specialists and radiologists if they cannot be interpreted. Identifying the Region of Interest (ROI) that leads decision-making to the network will enhance medical experts and data science experts' understanding.

The results provided by designed networks for the utilized datasets were investigated and explored. The Class Activation Mapping (CAM) [83] results were displayed to localize the areas questionable for the COVID-19 virus. To emphasize the distinctive regions, the probability, predicted by the DCNN model for each image class, gets mapped back to the last convolutional layer of the corresponding model that is peculiar to each class. The CAM for a determined image class is the outcome of the activation map of the Rectified Linear Unit (ReLU) layer following the last convolutional layer. That to what extent each activation mapping contributes to the final grade of that particular class is identified. The novelty of CAM is the total average pooling layer that is applied after the last convolutional layer, which is based on the spatial location to produce the connection weights; thereby, it permits to identify of desired regions in an X-ray image that differentiates the class specificity preceding the Softmax layer leading to better predictions. Illustrations in Fig. 15 and Fig. 16 using CAM for DCNN models allow the medical specialists and radiology expertise to localize the areas questionable for the COVID-19 virus. Figs. 15 and 16 indicate the results for COVID-19 detection in X-ray images. Fig. 15 shows the outcomes for the case marked as 'Covid19' by the radiologist, and the DCELM-ChOA model not only predicts the same result but also indicates the distinctive area for making a decision.

Fig. 16 shows the outcome for a 'normal' case in X-ray images. Different regions are emphasized by comparing both models for their predictions of the 'normal' subset. Now, medical specialists and radiology expertise can choose the network design based on these decisions. This type of CAD visualization introduces a secondary but useful opinion for the medical specialists and radiology experts to improve their understandings of deep learning models.

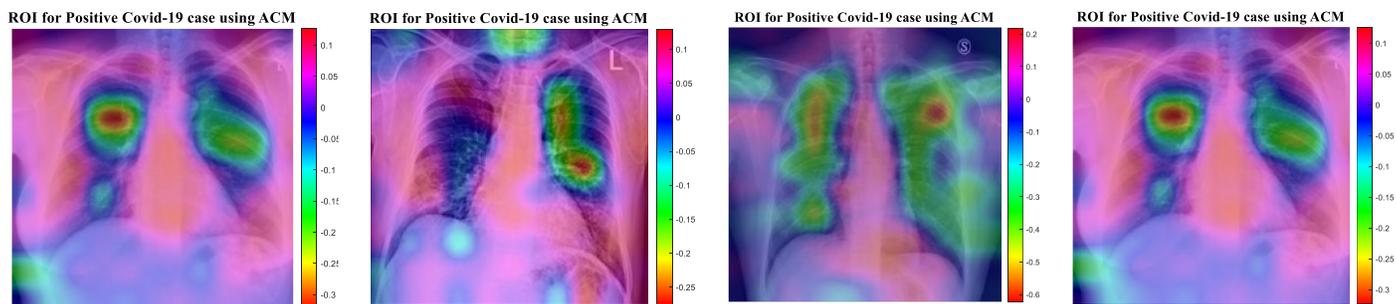

**Fig. 15** ROI for Positive COVID-19 Cases Using ACM

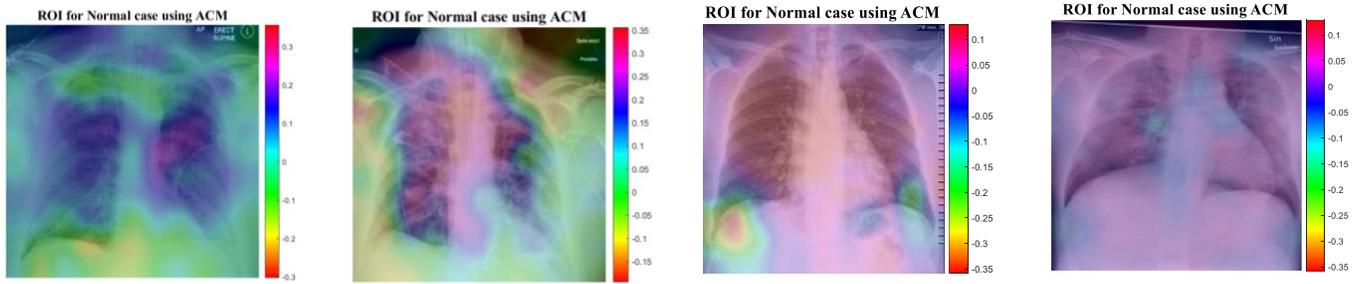

**Fig. 16** ROI for Normal Cases Using ACM

## 5. Conclusion

In this paper, the ChOA is proposed to design an accurate DCNN model for detecting positive COVID-19 X-ray. The designed detector was benchmarked on the *COVID-Xray-5k* dataset, and the results were evaluated by a comparative study with classic DCNN, DCELM, DCELM-GA, DCELM-CS, and DCELM-WOA. The results indicated that the designed detector can present very competitive outcomes as it is compared to mentioned benchmark models. The concept of Class Activation Map (CAM) was also applied to detect regions potentially infected by the virus. It was also found that it correlates with clinical results, as confirmed by an expert in advance. Limited lines of research direction can be proposed and introduced for future works with the DCELM-ChOA, such as detecting and classifying SONAR submarine targets; additionally, changing ChOA to tackle multi-purpose optimization problems can be recommended as another potential contribution. The investigation of the effectiveness of chaotic maps to improve the performance of the DCELM-ChOA can be another line of research direction. Although the results were promising, further studies are needed on a larger dataset of COVID-19 images for having a more comprehensive evaluation of the accuracy rates.